# Establishing, versus maintaining, brain function: a neuro-computational model of cortical reorganisation after injury to the immature brain

Sreedevi Varier[1] MRes; Marcus Kaiser[1,2,3] PhD; Rob Forsyth[2] PhD BM BCh

[1] School of Computing Science, Newcastle University, United Kingdom

[2] Institute of Neuroscience, Newcastle University, United Kingdom

[3] Department of Brain and Cognitive Sciences, Seoul National University, Korea

Correspondence to:

Dr R Forsyth

Institute of Neuroscience, Newcastle University,

Sir James Spence Building,

Royal Victoria Infirmary

Newcastle upon Tyne, NE1 4LP,

United Kingdom

r.j.forsyth@ncl.ac.uk

Word count 2835 (main text excluding Abstract) + 422 word Appendix

6 figures (all B&W) no tables

ABSTRACT

**Background and Purpose**

The effect of age at injury on outcome after acquired brain injury (ABI) has been the subject of much debate. Many argue that young brains are relatively tolerant of injury. A contrasting viewpoint due to Hebb argues that greater system integrity may be required for the initial establishment of a function than for preservation of an already-established function.

**Methods**

A neuro-computational model of cortical map formation was adapted to examine effects of focal and distributed injury at various stages of development. This neural network model requires a period of training during which it self-organises to establish cortical maps. Injuries were simulated by lesioning the model at various stages of this process and network function was monitored as "development" progressed to completion.

**Results**

Lesion effects are greater for larger, earlier, and more widespread lesions. The mature system is relatively robust, particularly to focal injury. Activities in recovering systems injured at an early stage show changes that emerge after an asymptomatic interval.

**Conclusions**

Early injuries cause qualitative changes in system behaviour that emerge after a delay during which the effects of the injury are latent. Functions that are incompletely established at the time of injury may be vulnerable particularly to multifocal injury.

**Introduction**

The effects of age at injury on outcomes after acquired brain injury (ABI) have been long debated. In some respects, young brains appear remarkably robust to injury. Early focal injury to dominant hemisphere language cortex need not result in aphasia [1] and functional hemispherectomy for treatment of intractable focal epilepsy can be tolerated [2], sometimes remarkably [3]. Such dramatic findings inevitably colour health professionals' views (as surveyed and criticised by Webb et al [4]) that it is "better to be injured young". In recent years however it has been recognized that this apparent resilience has a cost, and is not representative of all forms of brain injury at all stages of childhood [5, 6].

Addressing this issue clinically is challenging. A recent study [7] reported poorer outcomes from focal brain injuries acquired before the age of two. Subsequent debate over this paper however [8, 9] emphasised the difficulty of separating the timing of any acquired injury from its aetiology, size, extent (uni- or bi-hemispheric) and the presence of co-morbidities such as epilepsy. These factors are all to an extent age-dependent and thus threaten to confound age-at-injury studies. For example Lidzba et al [8] point out the greater prevalence of bi-hemispheric injuries in Anderson et al's younger children.

Contradictory messages about robustness or otherwise to early injury arise from literature studying different clinical populations. Studies of pre-, peri- and early postnatally-acquired focal insults, suggest milder cognitive deficits than would result from comparable insults in adults [10]. In contrast, deleterious effects of early injury that become increasingly evident with time are an emphasis of the traumatic brain injury (TBI) literature [11]. Whilst a very heterogeneous form of injury [12] TBI is characteristically widely distributed in a multifocal manner [13].

As well as aetiology-dependent variability in lesion size, number and location, the complexity of the age-at-injury literature results from interplay between age at injury, the time that has elapsed since injury, and the domain(s) of function under consideration. Age at injury must be related to milestones in the development of the neural structures supporting that function. In relation to the literature on motor function after very early injury for example the ontogeny of the crossed corticospinal projections to spinal alpha-motor neuron targets becomes relevant. This is well under way in the third trimester and complete by term but comprises both crossed and uncrossed projections [14]. It is followed by pruning of ipsilateral axonal projections through competitive inhibition during the first year or so of postnatal life and organisation of cortical representational maps. Extensive unihemispheric injury acquired pre- or early postnatally may be thus compensated by preservation of uncrossed projections from intact cortex [15], or even preservation of more wide-ranging peri-lesional cortico-subcortical projections that would normally regress [16].

Response to later injury however, once patterns of cortical projections to distant targets are established, predominantly involves reorganisation of cortical maps and here an important idea due to Hebb becomes relevant. On the basis of clinical observation of effects of paediatric frontal lobe injuries he hypothesised: "it appears…that an early injury may prevent the development of some intellectual capacities that an equally extensive injury, at maturity, would not have destroyed…Some types of behaviour that require a large amount of brain tissue for their first establishment can then persist when the amount of available tissue has been decreased…More fibres are necessary [for the first establishment of an assembly] than for its later function" [17].

The ideas informing Hebb's views are somewhat unintuitive, but potentially important to understanding how effects of injury in the slightly older child can become increasingly evident with time [11].

We aimed to examine and illustrate them using a neuro-computational model of brain injury. The need to specify explicitly the properties of such models provides insights into the characteristics that determine particular behaviours of interest, allowing assessment of their biological relevance and plausibility. Such a model also allows effects of lesion timing, size and extent to be investigated independently in a way that will never be possible clinically. Hebb's hypothesis about the differences between the first establishment and later maintenance of a function arises from insights into the importance of *intracortical* interactions in cortical reorganisation after injury. Such interactions may be overlooked in discussion of injury to the young brain by considerations of re-organisation of distant cortical projections, however in a child (as opposed to an infant) such processes are likely to be highly relevant.

The model used in this study was initially developed as a self-organising model of cortical map formation in a simple two-dimensional "somatosensory cortex" [18, 19]. This type of self-organising model requires a period of "training" during which initial random activation propagates recursively through the model. Two properties of the model lead to the spontaneous formation of cortical maps where neurons driving a given muscle tend to cluster. These properties are *(i)* a synaptic learning rule (often known as a Hebbian learning rule) where synaptic weights strengthen if pre- and post-synaptic activity are correlated in time and *(ii)* the implementation of "Mexican hat"-type lateral peri-stimulus inhibition (see Appendix). We hypothesised that cortical map formation was an important element of Hebb's concept of "establishment of an assembly of neurons to support a function". Such maps are not static and "hard wired", but form as the result of dynamic equilibria between inputs competing for targets. As a result these maps can be perturbed both by injury and rehabilitation after injury [20, 21]. Outside the period in the very young child in which re-direction or re-allocation of *distant* projections can be an important component of the response to injury (see above), changes in cortical maps reflecting redirections of *local* cortico-cortical projections are strong candidates as the neural substrate both for the impairments seen after injury and any rehabilitation-related recovery. Hence we chose to extend studies of the Reggia model to examine effects of injury in situations before the training and self-organisation phase had completed, allowing a direct examination of Hebb's views on the interaction between ongoing development and injury before map formation was complete (see also Discussion).

**Methods**

The model used in this study is adapted from that originally developed by Reggia and colleagues [18]. The model, of the neurological control of a simple upper limb, is described in more detail in the Appendix. The model is shown in Figure 1 and comprises four "layers" in a closed loop of interactions. A motor cortex (layer 1) determines the activation of three agonist-antagonist pairs of muscles (shoulder extensor/flexor, shoulder abductor/adductor, and elbow flexor/extensor) in layer 2. These determine the position of a virtual arm in 3D space and thus muscle "lengths" and "tensions" (layer 3). These data are fed back to proprioceptive cortex (layer 4), which connects back to the motor cortex thus closing the loop (Figure 1 and Appendix). Layers 1 and 4 are represented as 20 x 20 grids of hexagonally-tessellated neurons. Activity in the model is initiated by application of external activation that is then propagated through the model. The resultant change in the activity of neurons is governed by equations 1 and 2 (Appendix) that incorporate the influence of competition for activity received from any neuron. The inter-cortical weights, i.e. the strengths of all possible connections between one layer and the next, are allowed to self-adjust as a result of repeated application (several hundred times) of a Hebbian learning rule (Appendix). This has the effect of strengthening synapses with temporally-correlated pre- and post-synaptic activity, and weakening synapses with uncorrelated activity. During this period the network self-organises and forms cortical maps (Figure 2) emulating the self-organization observed in the cortex during development. Pilot studies showed that the intact network reliably stabilised after 800 cycles, which was taken to represent full "maturation". Convergence following lesion was empirically tested and confirmed to be between 800 and 900 cycles in all cases.

Lesions were introduced at 10, 20, 40, 60 or 80% of development (i.e. after between 80 and 640 cycles) and after lesioning, training was allowed to continue until 1000 cycles had completed to ensure full stabilisation. Computationally, a lesion was introduced by setting the activation in affected cells to zero and severing all associated incoming and outgoing connections (i.e. fixing the weights of all such connections at zero). (Uni-)focal lesions were implemented as contiguous rectangular patches of 1, 2, 4, 8 or 16% of the M1 motor cortex layer. Distributed, multifocal lesions were applied to 1-16% (i.e. between 4 and 64 cells) of the M1 layer cells at random. For each of the 25 combinations of lesion size and developmental stage, models were run five times to control for influences of different random initialisations. For each model an intact model with the same random initialisation was run as a control. For the contiguous lesions only, simulations were repeated placing them at five different locations within the motor cortex to examine whether lesion location (e.g. central location vs. an edge location) changed the findings qualitatively.

Simulations in all cases were continued beyond stabilization to 1000 cycles. To assess performance, muscle activity in each lesioned model was compared to the corresponding intact activity at every time step between 900 and 1000 cycles. The mean of this absolute percentage deviation over the 100 cycles was used as the quantitative comparison measure.

Results

*Intact model*

Based on the time-course of muscle activation (Figure 2 top) it was observed that 800 cycles of learning was sufficient for achieving self-organization. The muscle activation followed a logistic (sigmoid) learning curve and converged to the maximum value around this time. The organisation occurring is apparent from the cortical maps before and after the learning phase (Figure 2 B and C). Similar map organisation is observed for all six muscles.

*Lesions*

Rectangular lesions of 1-16% of the neurons in the motor cortex (4 to 64 elements) were introduced at 10-80% of "maturity" (80 to 640 cycles). The two parameters influencing the recovery are the size of lesion and its time of onset (Figure 3A-D). Figure 3A and 3B show effects of focal lesions; 3C and 3D show distributed injuries. For a given lesion size, injury effects are more marked (greater percentage deviation from intact) the earlier the lesion occurs and this is more obvious for distributed than focal injuries of the same size (compare figures 3A and 3C). The same effects are shown in figures 3B and D where data are organised by "age at injury". These finding were robust with respect to location of the lesion in the cortex for focal injuries.

The activation present in the muscles was plotted at each time step to observe changes at the local level (Figure 4). Muscle activities in recovering injured systems show a late-onset instability in activity not seen in the intact model. In the case of early lesions, this becomes obvious only at the later stages of development after a "latent phase". A similar pattern was observed with distributed lesions.

Figure 5 (A-E) depicts fluctuations in shoulder extensor activation relative to the intact model following 16% focal or distributed injury at varying stages of development. Distributed lesions (black) were more potent than focal lesions (gray) particularly when sustained early.

In later lesions, whether a lesion was focal or distributed mattered less in determining outcome (figure 5E). Similar results were observed for all sizes of lesions.

The comparative effects of distributed versus focal lesions of the same total size are shown in Figure 6. Each point represents a specific injury size (symbol size) and timing (symbol) combination. The mean percentage reduction in activation relative to the intact model for a focal injury determines the point's $x$ coordinate, and the corresponding value for distributed injury the $y$ coordinate. The fact that the majority of the points lie above the $x = y$ identity line demonstrates the greater effects of distributed injury.

**Discussion**

Greater functional deficits were seen with larger lesions and earlier injury. Although multi-focal and focal injuries yielded qualitatively similar results, deficits were most pronounced in early, distributed multifocal injuries. The other important general result of this work is that enduring qualitative changes in system behaviour compared to the intact model can emerge late after early injury, after a phase during which the effects of the injury are "latent". After early injury the system develops a late instability in muscle activation, which can take a few hundred cycles to emerge, not seen in the mature system at any time (Figure 4).

The aim of this project was to develop an *in silico* model of injury so as to better understand the properties that lead such a model to exhibit the vulnerability to early injury predicted by Hebb. He hypothesised that greater degrees of network integrity were required for the establishment of an assembly than might be necessary to maintain its later function. As emphasised in the introduction this arises because of the importance of intracortical interactions in driving reorganisation post injury. Any *structural* lesion of "dead neurons" creates an additional *functional* lesion in adjacent cortex which whilst intact, has lost connections from the lesion itself. The *functional* lesion is more widespread (more fibres are affected) in a distributed lesion than a comparable focal lesion and could account for the more potent effect of the former. Early in the training phase, all activations are near-zero. For a single neuron to strongly activate it needs the co-activation of its neighbours: an early lesion to a neighbour can prevent activation of a neuron. By contrast in the mature network, activation of a single neuron is much less dependent on the co-activation of a neighbour: indeed the strong lateral inhibition that develops in the model means that lesioning of a neighbour has little effect on a neuron's activation. In situations where two neurons are strongly connected, so that lesion of the neighbour does affect function, this is mitigated to an extent by the fact that the connectivity also leads to redundancy – both the neuron and its lesioned neighbour have the same targets.

As with any neurocomputational model, this is an abstraction of biological reality, and a clear understanding of what is and is not represented in the model is important. Specifically, the model *only represents the processes of formation (and re-organisation after injury) of cortical maps*. It does not model the establishment of axonal projections to distant targets, or their reorganisation in response to injury: the architecture of connections between layers of the model remains constant. This is therefore not a model of injury at very early ages before adult patterns of cortical projection to distant targets have established, and where the potential for reorganisation of distant projection patterns is an important determinant of functional deficit [22]. Such major re-allocations probably underlie the relocation of language that mitigates effects of early injury to dominant language cortex [3, 10, 23] and are certainly important in determining deficits after injury to motor areas [14, 15]. The potential for wholesale pathway redirection is much reduced after neurons have reached their normal destinations, when restoration of function depends more on emergence of novel (now de-afferented) targets and in particular, re-organisation of juxta-lesional cortical maps [20, 21].

The Reggia model was initially developed as a model of cortical map formation as a result of competition for targets [18, 19]. For this to be an appropriate model to the clinical question at hand a number of conditions need to be met. The first is that the processes reflected in cortical map perturbations after injury and rehabilitation are directly relevant to the clinical impairments and recoveries of function seen clinically (in whatever domain). That this is so, and that cortical maps are not fixed but result from dynamic equilibria that can be perturbed [20, 21] has been central to the recent renaissance of rehabilitation neuroscience. The second is that both normal development and recovery after injury can be represented by the same model. Although again clearly an oversimplification this is biologically plausible [24].

There are clearly important limitations to the extent to which the process of training the model should be regarded as equivalent to physiological development. Here network training is a process of intrinsic, undirected self-organisation whereas in physiological development (and indeed successful rehabilitation) it is exquisitely shaped by external stimulation. However the findings predicted by the model of greater vulnerability to early, distributed injury arise primarily from simple features of the network (such as the dependency on neighbours for co-activation) rather than the specifics of how the network is trained. The model was conceived as a representation of the motor system but the basic form of neuronal output (layer 1), feeding forward to an effector organ (layer 2), interacting with the environment and resulting in sensory feedback (layer 3) which is projected to sensory cortex (layer 4) could also be regarded as loosely representative of other systems such as language cortex. It is however a "single system" model that does not incorporate cross-modality cortico-cortical or important cerebello-cortical influences and cannot as yet model the processes of organisation of such projections with development currently being identified by resting state fMRI and other techniques [25, 26] . Additionally the insult is represented as confined to Layer 1. The model does however have the potential to study the effects of other injury-related properties. For example classical neurology has long recognised the importance of lesion "momentum" in determining the manifestations of injury: sudden-onset injury (e.g. from trauma or stroke) tends to be more clinically evident than slowly-developing dysfunction e.g. due to an expanding tumour. The lesions induced here are "high momentum" but gradually evolving lesions could be modelled. Additionally the effects of sequential injury [27] would be addressable.

The purpose of the model is to illustrate the contrasting effects of injury before, during and after the establishment of mature cortical organisation in a single "domain". In broad terms motor systems organise before language systems in human development, with important establishment of major cognitive functions not establishing until the second decade of life [28]. Thus the approximate biological age to which the model corresponds will depend on the domain in question, but supports the Hebbian understanding [17] of why particularly diffuse injury at a young age will particularly affect aspects of frontal lobe function which are most likely to be incompletely established at injury.

Neurocomputational models provide insights into emergent properties of complex systems. An additional benefit is that they require explicit statements of assumptions, approximations and simplifications and thus throw into sharp relief the validity of these assumptions. Inevitably they tend to model single aspects of biological reality. Nevertheless they have provided biologically plausible insights into effects of injury and rehabilitation [29]. An important aspect of development not implemented in this model is the phenomenon of sensitive periods or "critical windows" [30] which have also been modelled computationally in various ways {Thomas [31, 32]. Incorporating such considerations into this model would essentially comprise changing the model's properties over time. Our model does contain a "learning constant" (see Appendix): incorporating a progressive decay in this parameter (to simulate decreasing "general plasticity" with age) increases model stabilisation times but does not alter the qualitative findings of this study (data not shown).

Addressing the age at injury question empirically through clinical studies is complicated by age-dependent differences in injury aetiology and mechanism. *In silico* models have the potential to inform this debate and generate new hypotheses. This work highlights the need to distinguish very early injury (where reallocation of cortical projections are important influences on manifestations of injury) from injury in the slightly older but still immature child, where (as in the adult) responses to injury are largely constrained to adjustment of cortical maps. For reasons set out above this is a model of injuries acquired in slightly later childhood (such as TBI) rather than pre- or peri-natally acquired injury. Findings of relative vulnerability of the immature brain to distributed injury, and latent injury effects, are particularly consistent with the clinical literature surrounding TBI [33] and brain tumours in young children [34], where the poorer outcomes seen in children relative to adults is often couched in terms of the balance between greater plasticity in the immature nervous system, versus the challenge of having to complete development to maturity ("make a year's progress every year") with an injured brain.

**Acknowledgements**

M.K. was supported by WCU program through the National Research Foundation of Korea funded by the Ministry of Education, Science and Technology (R32-10142), the Royal Society (RG/2006/R2) and the CARMEN e-science project (www.carmen.org.uk) funded by EPSRC (EP/E002331/1). M.K. and S.V. were supported by EPSRC (EP/G03950X/1).

**Conflict of Interest**

No conflict of interest.

LEGENDS TO FIGURES

*Figure 1*

Schematic representation of neurocomputational model. See Methods, Appendix and references [18] and [35] for further information.

*Figure 2*

Maturation of intact system. (A) Development of activity in shoulder extensor muscle stabilises within 800 training cycles (other muscles show similar patterns). (B-C). Establishment of cortical maps. Dots show neurons with connection weights to the shoulder extensor muscle greater than a threshold of 0.15. In the immature system (cycle zero, bottom left) connectivity to the extensor is distributed. After training (cycle 800, bottom right) connectivity has organised into discrete areas.

*Figure 3*

Lesion effect expressed as mean percentage deviation in activity at the end of learning relative to intact model, plotted against age at injury (A and C) and lesion size (B and D). Focal lesions are shown in A and B; distributed lesions in C and D. Note that plots A and B have a larger Y-axis scale than plots C and D to aid visualization.

*Figure 4*

Time-course of muscle activation before and after a 16% focal lesion introduced at various stages (represented by the vertical dashed line)

*Figure 5*

Percentage deviation in activity of Extensor after lesion-onset, from intact, followed along time-course of learning for a 16% focal (grey) or distributed (black) lesion introduced after 80 (A), 160 (B), 320 (C), 480 (D) or 640 (E) cycles.

*Figure 6*

Comparison of distributed versus focal injuries. Each point represents a specific injury size (symbol size) and onset age (symbol) combination. The mean percentage deviation in activity (after stabilization) relative to the intact model for a focal injury determines the point's *x* coordinate, and the corresponding value for distributed injury the *y* coordinate.

**APPENDIX: NEUROCOMPUTATIONAL MODEL**

**The computational model**

The computational model consists of a four-layered, closed loop neural network (Figure 1). There are two cortical layers, denoted M1 and P1, representing the primary motor cortex and the proprioceptive cortex respectively. Neurons within these cortical layers are hexagonally connected to six neighbours in the same layer within a radius of 1. Edge neurons are connected to their counterparts on the opposite edge to form a torus and to avoid edge effects. Additionally, each element in P1 has a loose topographic connection to its corresponding element in M1 and the latter's neighbours within a radius of 4. The motor layer holds the activations of each of the six muscles and the proprioceptive input layer stores the corresponding length and tension of the muscles, representing the position of the arm. Activation flows sequentially through all the layers forming a closed, self-organizing loop. The change in activation in each layer is computed using the following equations (details in [18]:

$$\frac{da_k}{dt} = [Ca_k] + [M - a_k][in_k + ext_k] \quad (1)$$

$$in_k = \sum_j out_{kj} = \sum_j c_p \frac{(a_k^p + q)w_{kj}}{\sum_l (a_l^p + q)w_{lj}} a_j \quad (2)$$

$$\Delta w_{kj} = \eta(a_j - w_{kj})a_k \quad (3)$$

where, $a_k = a_k - \alpha$ if $a_k > \alpha$ otherwise $a_k = 0$ \quad (4)

$a_k$ is the activation of element $k$ at a given time $t$. The rate of change in $a_k$ (i.e. d$(a_k)$/dt) is dependent on the present value of $a_k$ multiplied by a decay constant $C$, plus the sum of the total activation received by element $k$ from all other elements ($in_k$) and any external driving stimulus $ext_k$ (collectively multiplied by a fixed scaling factor $M\text{-}a_k$). The external stimulus $ext_k$ is zero except for layer M1 where a "patch" of activation was supplied to a randomly selected neuron representing activation from other areas in the brain. The input to element $k$ ($in_k$) equals the sum of the outputs from its afferents ($\sum_j out_{kj}$) given by eqn *(2)*. $C_p$ is an output gain constant. Parameters $p$ and $q$ result in a biologically-inspired "Mexican hat" pattern of peri-stimulus inhibition. Synaptic weights ($w_{kj}$) are adjusted according to a learning rule *(3)* where $\eta$ is the learning constant referred to in the text. This equation implements a Hebbian learning paradigm where synaptic weights increase if pre-synaptic and post-synaptic activity are correlated and decrease otherwise. Equation *(4)* ensures that only substantially activated elements (where $a_k$ exceeds a threshold α) can learn.

The model was initialized with random inter-cortical weights (between 0.1 and 1) and activation was set to zero in all the layers. The activation in each layer was allowed to stabilize over 120 iterations and then learning was applied to the inter-cortical weights by adjusting weights using the equation *3* rule. The intra-cortical weights remained constant throughout learning.

Figure 1

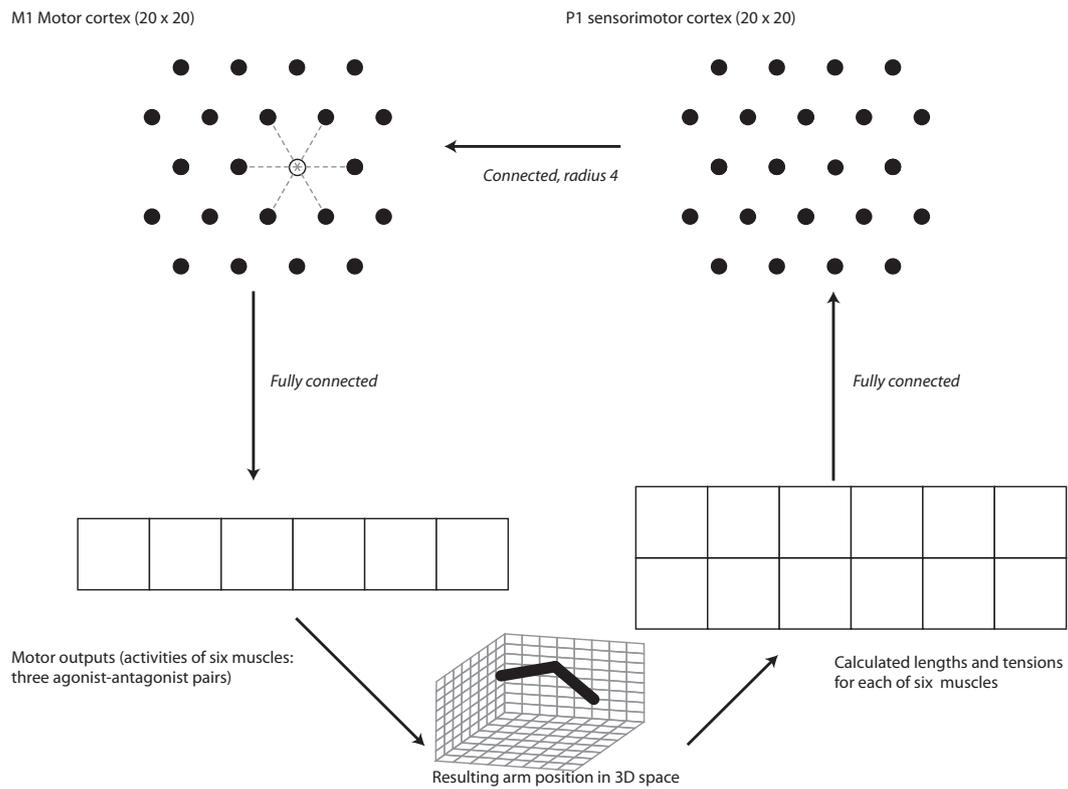

Figure 1

Figure 2

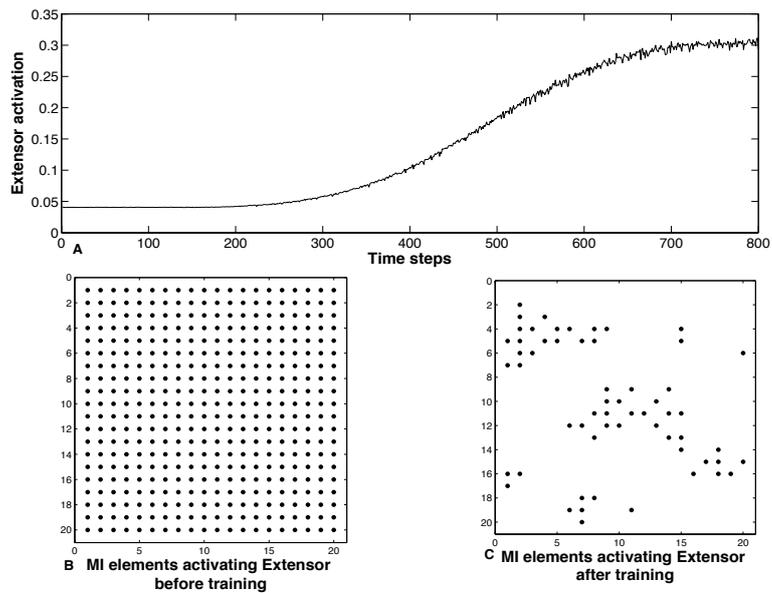

Figure 3

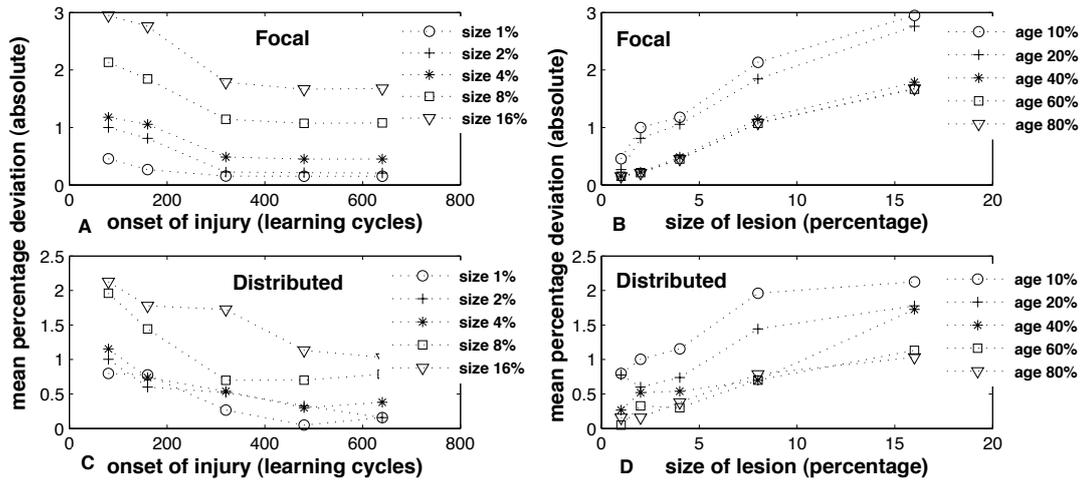

Figure 4

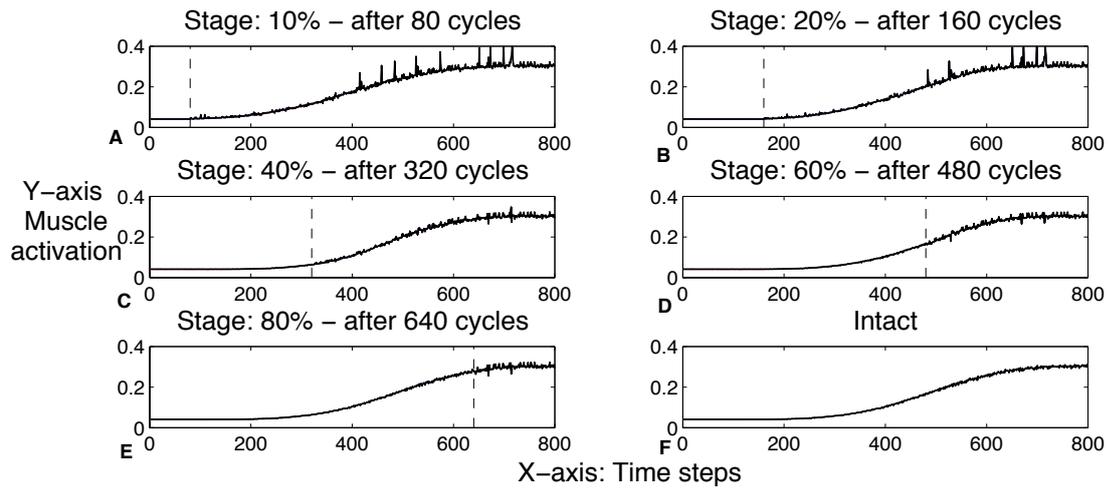

Figure 5

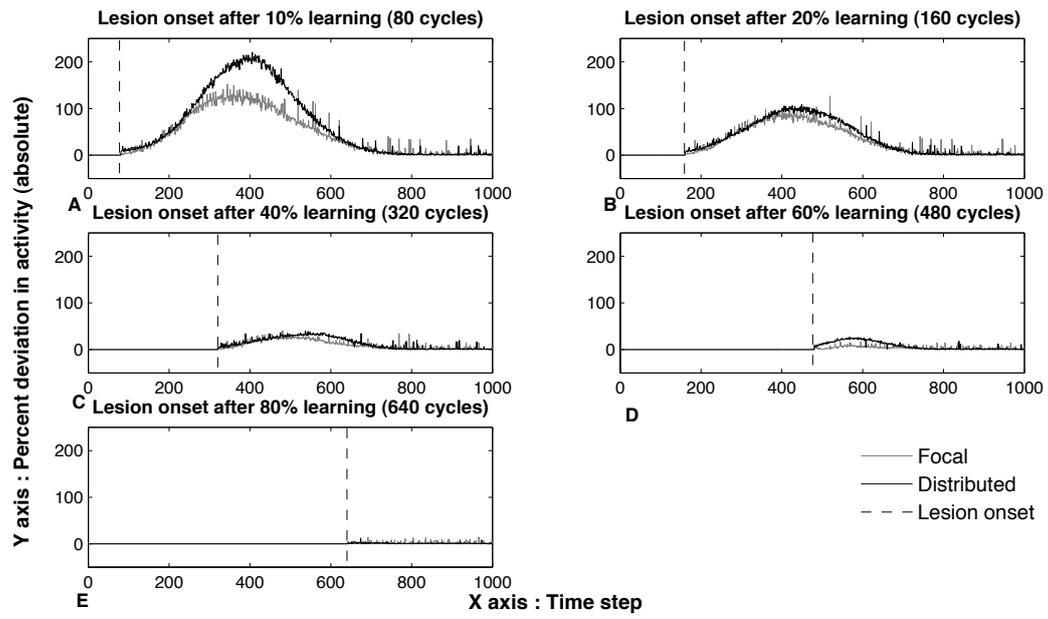

Figure 6

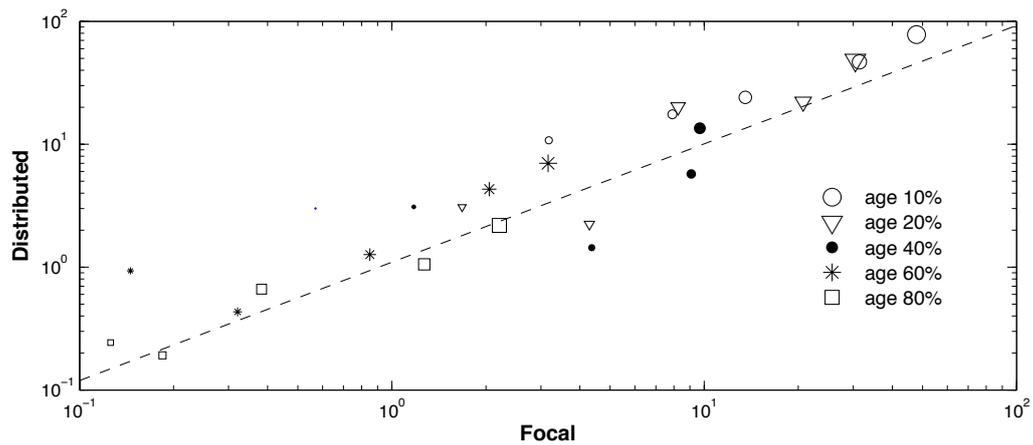